# The fluid running in the subnanochannel with functional surface

Hui Xie*, Kunyi Xie

(Key Laboratory of Low-grade Energy Utilization Technologies and Systems of Ministry of Education, College of Power Engineering, Chongqing University, Chongqing, 400030, China) [1]

# Abstract

**We have researched the motion of gas in the subnanochannel with functional surface which wettability has a gradient for the fluid by using molecular dynamics simulation. The results show that the gas is driven to flow under a single heat source and without any other work or energy applied to the system. The driving source is owed to the potential gradient of the functional face which keeps the fluid running in the subnanochannel. The width of the channel and the pressure of the reservoir has a significant influence on the flow velocity, which, respectively, has an optimal value for the maximum velocity. The flow velocity grows with the increasing temperature.**

**Key words: nanochannel; molecular dynamics; functional surface; flow**

**Introduction**

Because the interaction between atoms becomes the dominant effect in the nanoscale, the flow in the nanochannel is different from the macroscopic. The molecular dynamics simulation becomes a powerful method to research the flow in the nanoscale. There are some wall characteristics influencing on the nanoscale flow, such as the stiffness[1, 2], the lattice direction[3, 4] and the wettability[5]. The driving forces, used in the literatures, include the pressure[6, 7], the temperature gradient[8] and the gravity[5, 9]. Due to the wall effect, the velocity slip and density stratification are typical phenomenon of the flow in nanoscale[10, 11].

The second law of thermodynamics is widely accepted and is initially constructed on the experience of the macroscopic for human beings[12]. Is there situation for smaller scale that the second law is unkept? Alhambra et al[13] pointed out that the probability of thermodynamical transition without work can be finite for finite-sized and quantum systems. Lesovik et al[14] also uncovered that the second law can be violated for some special situations of quantum system. We report a flow in the subnanoscale driven with a single heat source researched by using molecular dynamics simulation. The driving source is owed to the potential gradient of the functional face which keeps the fluid running in the subnanochannel. The width of the channel and the pressure of the reservoir has significant influence on the flow velocity, which, respectively, have an optimal value for the maximum velocity. The flow velocity grows with the increasing temperature.

*This work is supported by the National Natural Science Foundation of China (Grant No. 51206195), Natural Science Foundation of Chongqing (Grant No. cstc2013jcyjA90009) and the Fundamental Research Funds for the Central Universities of Ministry of Education of China (Grant No. CDJZR12110033).

†Corresponding author.   Email: xiehui@cqu.edu.cn ; Tel:   .023-65102469

**Results and disscussion**

The simulation system is shown in the figure 1(a). The system only exchanges heat with a heat bath. The subnanochannel with functional surface is connected with two reservoirs with the same pressure at the end. The h is used in the Lennard-Jones potential function to adjust the interaction between the fluid and the wall atoms that the functional surface, which wettability is not uniform for the fluid, is constructed. The atoms passed from inlet to outlet through the subnanochannel is defined as the positive flux (PosF) and the opposite is the negative flux (NegF). The difference between the PosF and NegF is the net flux (NetF). As shown in the figure 1(b), because of the thermal motion, the PosF and NegF both increase with the time and the PosF is always more than the NegF. So, the NetF through the subnanochannel is created, and it also increases with the time. That means the fluid is driven to flow. In the initial stage, there is a gradual acceleration of flow with time. After about 0.3 µs, the NetF increases linearly with time. From the slope of the NetF varying with the time (from 0.5 µs to 2.0 µs) and the density of the fluid in the subnanochannel, the average flow velocity can be calculated, which is 0.151 m/s for the condition of figure 1(b). In contrast, figure 1(c) shows the flux in the subnanochannel with nonfunctional surface, which means that the h is always equal to 1, under the same conditions of figure 1(b). It is obvious that there is only some thermal fluctuation and no flow.

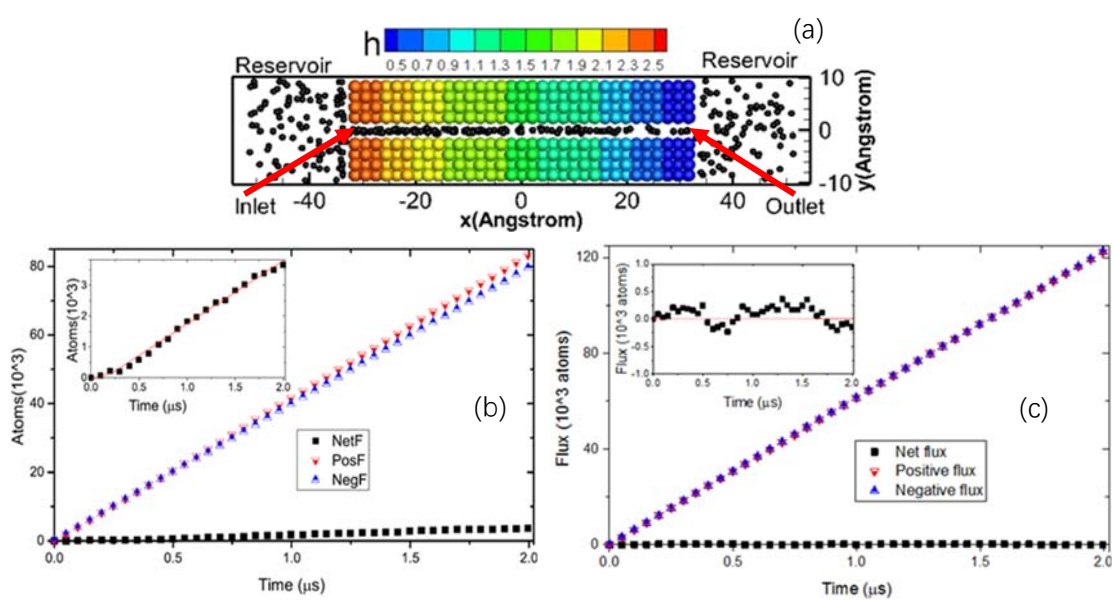

Figure 1 (a) The front view of the simulation system for W=5.5 Å (the width of the subnanochannel). The atoms varying with time through the functional (b) and nonfunctional (c) subnanochannel for W=5.5 Å, T=300 K (the temperature of the heat bath) and P=11.2 MPa (the pressure of the reservoir).

The next question is how the fluid is driven. Because there is no flow if the subnanochannel does not have the functional surface, as shown in figure 1(c), the driving source must come from the interaction between the atoms of the fluid and the functional wall. Barisik et al[15] found that the gas–wall interactions is the dominating factor for the transport in the near-wall region. Figure 2(a) and (b) shows the density distribution of the fluid in the subnanochannel for the condition in figure 1(b). The stratification of fluid is a typical phenomenon in nanoscale due to the wall effects[7, 16] and there is a high density domain close to the hydrophilic surface and a low density domain close to the hydrophobic surface[5]. The fluid atoms in the subnanochannel will 'feel' the different potential of the functional surface. As shown in Figure 2 (c) and (d), It is obvious that there are local concaves and global gradient of the potential. From the comparison of figure 2(a) and (c), (b) and (d), for the local, it is can be found that the fluid atoms 'would like to stay' in the concave. Sofos et al found the similar behavior that the fluid atoms would be trapped in the grooves region with a significant time[17, 18]. The fluid atoms would randomly get enough energy, through collision with other atoms including the wall and the other fluid atoms, to firstly overcome the lower energy barriers. That means the fluid atoms have the higher probability of moving from left to right than the opposite. So, the fluid would be continual flowing in the subnanochannel as the global potential gradient exists. The interlaced potential ratchets between upper and lower walls make the fluid atoms easily move from one concave to another. Liu et al[8] found that the flow in nanochannel can be driven by the wall potential gradient caused by the temperature gradient, which is kept by source and sink of heat. For the subnanochannel with functional surface, the flow can be maintained by a single heat source. It is known from the second law of thermodynamics that the heat from a single heat source cannot be extracted and converted directly to work. However, because there are only limited number of atoms confined in the subnanochannel with functional surface, which induces potential and density gradients, the equilibrium state, maybe, would never be reached. There is thermal motion, so some random atoms of the fluid would always get enough energy through collision with others to preferentially overcome lower energy barriers, moving to the potential gradient direction of higher to lower. That is to say, the unordered thermal motion of wall atoms is converted to the ordered flow of the fluid atoms by the potential gradient of the wall.

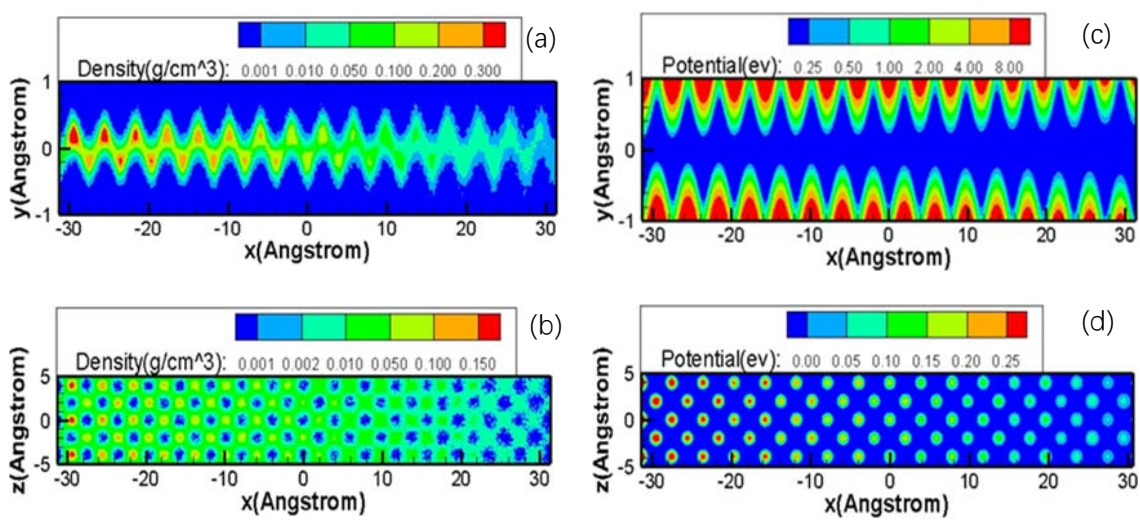

Figure 2. Two-dimensional contour of the density distribution of the fluid in the subnanochannel of x-y plane at z=0 (a) and x-z plane at y=0.2 (b) for the condition in figure 1(b); The potential distribution of x-y plane at z=0 (c) and x-z plane at y=0.2 (d) for W=5.5 Å.

Figure 3(a) shows the flow velocity varies with the width of the channel for T = 300 K and P = 11.2 MPa. For the small width, such as for W = 4.9 Å, there is no atoms in the channel, certainly, so the flow velocity is zero. That indicates that the obstacle of the inlet keeps the atoms out of the subnanochannel and is the main resistance for the flow in the narrow subnanochannel. As the W increases, the resistance of the inlet and the shear viscosity[19] decrease and the flow velocity increases and obtains a maximum value when the W is 5.5 Å. After that, the flow velocity decreases with the increasing W. That is because more and more atoms will 'find' their equilibrium site with the W increasing. That means that the driving 'force' reducing, instead of the resistance of the inlet, becomes the reason for the flow velocity drop. When the W is 19.1 Å, the flow velocity is close to the zero. Figure 3(b) shows the flow velocity as a function of the pressure of the reservoirs (P) for T=300 K and W=5.5 Å. It is obvious that there is an optimal P for the flow. For the lower pressure, similarly with the narrow subnanochannel, there is less atoms for the subnanochannel to drive. With the increasing P, the flow velocity gets the highest value when P=6.4 MPa. For the higher pressure, the viscosity of the gas increases with increasing pressure that induces the flow velocity falling. In figure 3(c), the flow velocity grows with the increasing temperature from 300 K to 900 K. Because the collision between fluid and wall atoms is the driving source and the collision is more and more violent with the increasing temperature, the fluid atoms get higher speed and more possibility moving to the potential gradient direction.

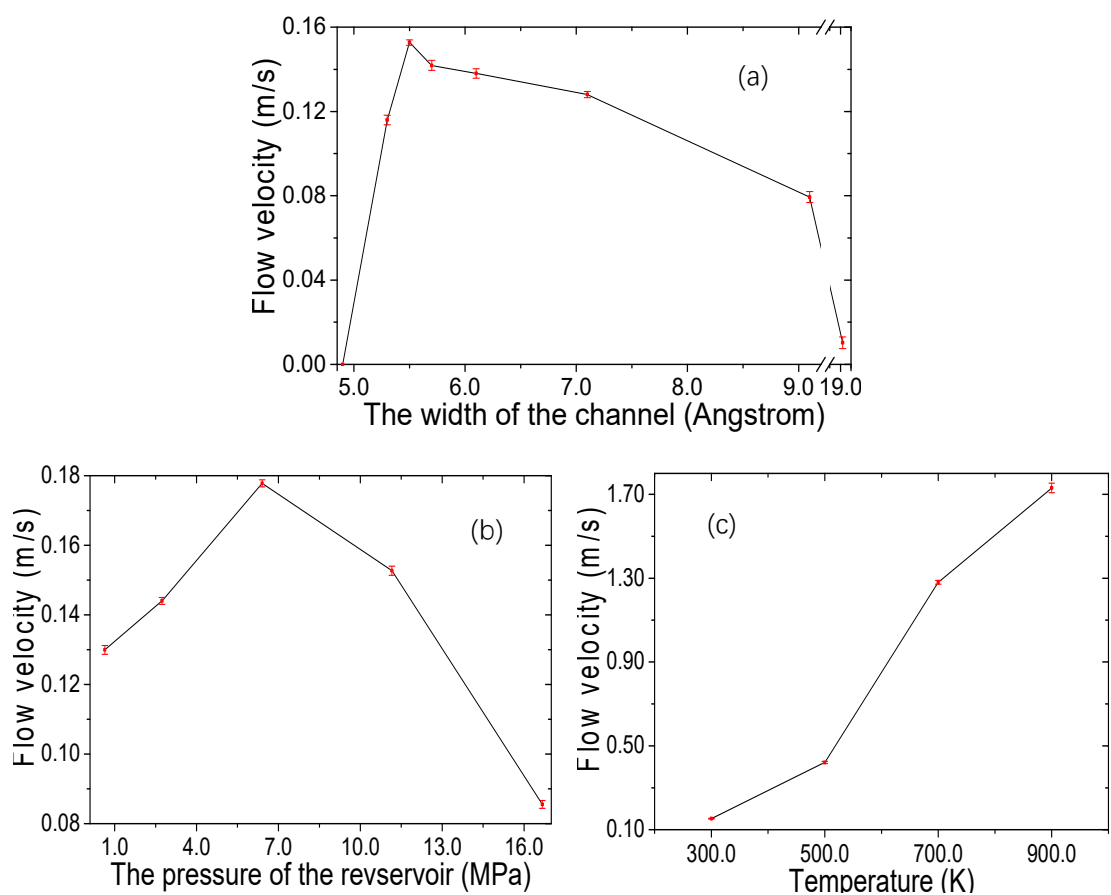

Figure 3. The flow velocity as a function of (a) width for T=300 K and P=11.2 MPa, (b) pressure for T=300 K and W=5.5 Å and (c) temperature for P= 11.2 MPa and W= 5.5 Å.

The spring constant, K, called stiffness, is a key parameter that describe the materials of the wall model used in molecular dynamics[1]. So, it is artificially changed in a broad range[1, 2] to study its influence on the flow. Figure 4 shows the flow velocity as a function of the stiffness when T is 300K, P is 11.2 MPa and W is 5.5 Å. When the stiffness is the half of the platinum, the fluid velocity hits the highest value. Under the same temperature, the average displacement of the wall atoms relative to their equilibrium position is increased at smaller value of K, which indicate the wall roughness increasing[2] that leads to the flow velocity decreasing. For the higher K, the energy of the fluid atoms getting from collision with the wall atoms will decline, that results in the flow velocity reduction. In short, the stiffness has influence on the flow velocity, but do not 'deny' the fact that the fluid is driven to flow in the subnanochannel by the potential gradient under a single heat source.

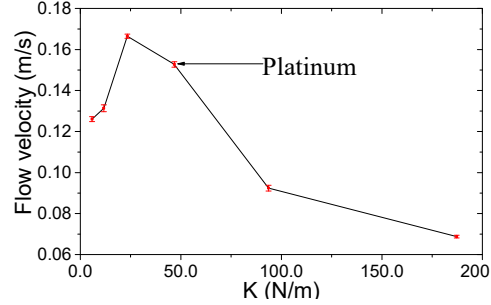

Figure 4. The flow velocity as a function of the stiffness of the wall for T=300 K, P= 11.2 MPa and W= 5.5 Å.

**Conclusion**

In a summary, the fluid confined in the subnanochannel with the wettability gradient surface can be driven by the potential gradient, maintained with a single heat source. The flow velocity is signally influenced by the width of the channel and the pressure of the reservoirs. The increasing temperature can accelerate the flow. The stiffness of the wall has quantitative influence on the flow velocity, but do not qualitatively change the continual flow. Although we have not studied these factors in any detail, the diameter of the fluid atoms, the number of the layers of the potential gradient, the width of every layer, the variation of the wettability gradient, and their couple effects. These factors should be considered in the future research. Maybe, faster or larger scale flow can be found.

**Method**

A simple fluid commonly used in molecular dynamics, gas argon[5, 11], is employed as the working fluid and platinum-like[5] is employed as planner wall. The computational cell size is 102.8×Y×39.2Å$^3$, where Y depending on the width of subnanochannel (W) (distance between the innermost atomic layers of the two walls) is changed from 18.7 to 32.9 Å, correspondingly, the W is from 4.9 to 19.1 Å. One wall is constructed with 1320 platinum particles arranged as a fcc lattice with 33, 4, 21 layers in the x, y, z direction, and another wall is constructed by 'moving' the first wall in the y direction, the moving distance depending on the W. The (010) surface is contacting with the fluid. The reservoir length in the x direction is 20 Å. The periodic boundary condition (PBC) is applied in the x, y and z directions, and the PBC in x direction can keep the two reservoirs with the same pressure. The numbers of argon atoms, depending on the different temperature, pressure and the width of the channel, varies from 33 to 430. The argon-argon and argon-platinum interaction is calculated with Lennard-Jones 12-6 function, $u(r) = 4\epsilon[(\sigma/r)^{12} - (\sigma/r)^6]$, the parameters $\epsilon$ and $\sigma$ are, respectively, the energy and length scales of the interaction, and $r$ is the distance between interaction atoms. For argon-argon interactions, $\epsilon = \epsilon_{Ar-Ar} = 1.67 \times 10^{-21} J$, $\sigma = \sigma_{Ar-Ar} = 3.405 \times 10^{-10} m$. For argon-platinum interactions, $\epsilon = h\epsilon_{Ar-Ar}$, $\sigma = \sigma_{Ar-Pt} = 3.085 \times 10^{-10} m$. The h is used to adjust the interaction between argon and platinum atoms indicating different wettability, as used in literatures[20-22]. The wettability can be controlled by chemical composition. In this letter, for the functional surface, the h is changed from 2.5 to 0.5 with arithmetic sequence as shown in figure 1(a) for 11 floors, each floor is consistent with 3 layers of platinum-like atoms, 1 fcc lattice, in the x direction. The spring potential is used for the platinum-like atoms, and the spring force constant K is 46.8 N/m[23, 24]. The cutoff distance is $2.5\sigma_{Ar-Ar}$.

The simulations were performed using large-scale atomic/molecular massively parallel simulator (LAMMPS)[25]. The NVT ensemble is used with Nose-Hoover thermostat with the same temperature for fluid and wall atoms, which can be seen as a single heat source. The relax time of the thermostat is 100 fs. In order to avoid the thermostat influence on the flow speed, it only be used to the y, z direction for the fluid atoms[10, 26]. The timestep is 1 fs. 10 ns are used for relax when the bulk speeds of fluid at x, y, z directions are adjusted to zero. 2 microseconds are used to calculate the fluid properties. The first 0.5 μs is seen as the developing time for the flow and the last 1.5 μs is used to calculate the flow velocity. The flow velocity, contrasted with the fluid atoms speed of thermal motion, is so small that it cannot be calculated by the time average as used in literatures[5, 23], so the positions of fluid atoms are monitored to calculate the number of the fluid atoms passed through the subnanochannel.